# The coupling of polarization and oxygen vacancy migration in ferroelectric $Hf_{0.5}Zr_{0.5}O_2$ thin films enables electrically controlled thermal memories above room temperature.

Dídac Barneo[1,2], Rafael Ramos[3], Hugo Romero[4], Víctor Leborán[4], Noa Varela-Domínguez[3], José A. Pardo[4,5], Francisco Rivadulla[3], Eric Langenberg[1,2,*]

[1]Departament de Física de la Matèria Condensada, Universitat de Barcelona, 08028 Barcelona, Spain.
[2]Institut de Nanociència i Nanotecnologia (IN²UB), Universitat de Barcelona, 08028 Barcelona, Spain.
[3]Centro Singular en Química Biolóxica e Materiais Moleculares (CiQUS), Universidade de Santiago de Compostela, 15782 Santiago de Compostela, Spain.
[4]Instituto de Nanociencia y Materiales de Aragón (INMA), CSIC-Universidad de Zaragoza, 50009 Zaragoza, Spain.
[5]Departamento de Ciencia y Tecnología de Materiales y Fluidos, Universidad de Zaragoza, 50018 Zaragoza, Spain.

**Abstract. -**

Here we investigate epitaxial $Hf_{0.5}Zr_{0.5}O_2$ ferroelectric thin films as potential candidates to be used as non-volatile electric-field-modulated thermal memories. The electric-field dependence of the thermal conductivity of metal/$Hf_{0.5}Zr_{0.5}O_2$/YSZ devices is found to be hysteretic–resembling the polarization vs electric field hysteresis loops–, being maximum (minimum) at large applied positive (negative) voltages from the top metallic electrode. This dynamic thermal response is compatible with the coupling between the ferroelectric polarization and the oxygen ion migration, in which the oxygen vacancies are the main phonon scattering sources and the polarization acts as an electrically active ion migration barrier that creates the hysteresis. This new mechanism enables two non-volatile thermal states: high (ON) and low (OFF) thermal conductivity, with an ON/OFF ratio of 1.6. Both the ON and OFF states exhibit high stability over time, though the switching speed is limited by ion mobility in the YSZ electrode.

The finding of materials that allow tuning their thermal conductivity ($\kappa$) at will using an external stimulus (such as electrical,[1–9] magnetic,[10–13] optical,[14] thermal,[3] strain,[15] or electrochemical[16–20]) is key to store and process information using thermal currents,[21–24] instead of conventional electrical currents. This thermal-based technology has a clear potential to excel in terms of energy consumption, reuse, and recovery,[24] as heat would no longer be a residual waste byproduct as it occurs in many industrial processes and electronic devices.[25] Yet for these materials to be used as binary thermal memories–an essential element in any information technology–, several requirements need to be met:[23] *i*) sufficiently large ratio between the high thermal conductivity state ($\kappa_{ON}$) and the low thermal conductivity state ($\kappa_{OFF}$), otherwise the binary nature becomes blur and ineffective; *ii*) rapid transitions (in seconds or less) from $\kappa_{ON}$ to $\kappa_{OFF}$ (and vice versa) using the external stimulus, to become competitive; *iii*) both the materials and the mechanism for the external stimulus must be easy to implement in a solid-state structure device; *iv*) robustness of the $\kappa_{ON}$ and $\kappa_{OFF}$ states (*i.e.* stable values along time once the external stimulus is removed), and *v*) endurance in the $\kappa_{ON}$–$\kappa_{OFF}$ switching cycles.

Among all the previously mentioned trigger mechanisms to alternate between the $\kappa_{ON}$ and $\kappa_{OFF}$ states, the electric field is the easiest one to be integrated in a solid-state thermal memory–basically just two electrodes and an applied voltage are needed–. In addition, the electric field can be applied locally, allowing a high density of "thermal bits". Moreover, low voltage values can generate a sufficiently large electric field, especially in thin-film solid-state architectures, reducing the energy consumption.[26] In this regard, ferroelectric materials have played a pivotal role in electrically tuning $\kappa$ thus far using several approaches: *i*) by electrically modifying the density of domain walls behaving as active phonon scattering centers in $PbZr_{1-x}Ti_xO_3$ or $BaTiO_3$ polycrystalline samples, and in PMN-PT single crystals;[1,2,4,9] *ii*) by electrically inducing a phase transition from antiferroelectric to ferroelectric in $PbZrO_3$ epitaxial films–each phase with different $\kappa$–;[3] *iii*) by electrically switching between two types of ferroelectric-ferroelastic domains–a- and c-domain displaying different $\kappa$–in single crystal $BaTiO_3$;[8] and *iv*) by electrically changing the phonon dispersion and, thus, indirectly $\kappa$ through the phonon velocity in single crystal $PbZr_{1-x}Ti_xO_3$.[7] The main advantage of using ferroelectrics, particularly in electrically modulating the domain wall density, is the rapid $\kappa_{ON}$–$\kappa_{OFF}$ switching, which tends to be in the ns scale.[23] However the $\kappa_{ON}/\kappa_{OFF}$ ratio is in general quite low, ranging from 1.1 to 1.2,[1–4,7,9] except for the alternation between the a- and c-domain, in which 1.6 ratio between the κ of the a-domain and the κ of c-domain is reported.[8] In this latter case, though, the distribution and the sizes of the a- and c-domains are quite inhomogeneous throughout the bulk BaTiO$_3$ single crystal[8] to use these individual domains as a regular array of "thermal bits". Moreover, after electrically poling the pristine ferroelastic pattern (i.e. the distribution and sizes of the a- and c-domains), when the electric field is removed the resulted ferroelastic pattern significant differs from the pristine one,[8] limiting the reliability of the distribution of the "thermal bits".

One of the approaches that have achieved much larger κ$_{ON}$/κ$_{OFF}$ ratios (up to 5.4) is by perovskite (x=3)–brownmillerite (x=2.5) topotactic phase transformation via oxygen ion migration in epitaxial SrCoO$_x$, La$_{0.5}$Sr$_{0.5}$CoO$_x$, and (Ca,Sr)FeO$_x$ films.[16–18,20] The brownmillerite structure displays much lower κ than the perovskite structure due to the large amount of oxygen vacancies acting as point defect phonon scattering centers.[17] In fact, both cation and anion defects are extremely efficient in reducing κ in oxide materials.[27–30] Moreover, this oxygen migration can be triggered electrically at 280ºC using yttria-stabilized zirconia (YSZ)–Y$_2$O$_3$:ZrO$_2$–substrates as solid electrolyte,[17,20] without the need of ionic liquids[16,18,19] that hamper their integration in solid-state structures. Furthermore, the endurance of the perovskite-brownmillerite phase transition cyclability is substantially improved by the solid electrolyte compared to the use of ionic liquids or oxidizing gas atmospheres, which tends to damage the structure after a couple of cycles.[20] However, the perovskite-brownmillerite phase transition takes several minutes,[16–18,20,23] which entails a much slower κ$_{ON}$–κ$_{OFF}$ switching time than previous approach using electrically mobile domain walls in ferroelectrics. Recently, voltage-biased atomic force microscopy tip was used to locally dragged oxygen vacancies in SrFeO$_{3-x}$, La$_{0.6}$Sr$_{0.4}$CoO$_{3-x}$ and La$_{0.7}$Sr$_{0.3}$MnO$_3$ epitaxial films on SrTiO$_3$ substrates at room temperature, locally writing non-volatile low thermal conductivity states.[31] Yet to switch back to the as-grown thermal conductivity state (i.e. to erase the low thermal conductivity states) temperature, instead of voltage, was required, inhibiting a full electric field control of the κ$_{ON}$–κ$_{OFF}$ switching.[31]

Here, we propose the use of ferroelectric Hf$_{1-x}$Zr$_x$O$_2$ as the electrically active thermal barrier, to take advantage of the close relationship between the stabilization of the metastable polar phase and the local concentration of oxygen vacancies.[32–34] On the one hand, ferroelectric Hf$_{1-x}$Zr$_x$O$_2$ intrinsically contains anion defects.[16–18] On the other hand, the oxygen vacancy migration through epitaxial ferroelectric $Hf_{0.5}Zr_{0.5}O_2$ (HZO) films grown on $La_{0.67}Sr_{0.33}MnO_3$/Nb:$SrTiO_3$ and using non-reactive metals like Au as top electrode has been proved to be quite fast (in seconds).[35]

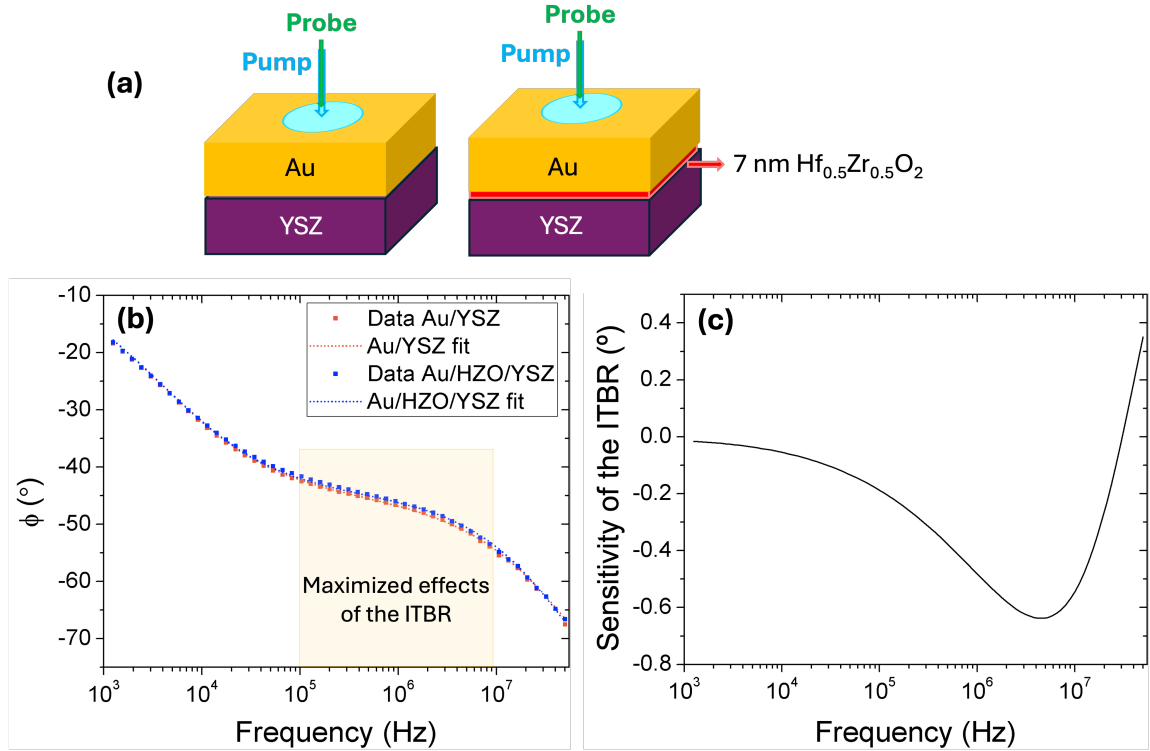

*Fig. 1. (a) Sketch of the FDTR measurements of the Au/$Hf_{0.5}Zr_{0.5}O_2$/YSZ and Au/YSZ samples. $Hf_{0.5}Zr_{0.5}O_2$ is modelled as an interface thermal boundance conductance as the Au/YSZ interface. (b) Frequency dependent phase (ϕ) data of the FDTR measurements. The frequency range where the effects of the presence or absence of the $Hf_{0.5}Zr_{0.5}O_2$ film are observed is shadowed in yellow. (c) Sensitivity analysis of interfacial thermal boundary resistance (ITBR).*

To assess this potential functionality of ferroelectric HZO as a thermal memory, epitaxial HZO films, 7 nm thick, were grown on isomorphic YSZ substrates, (111)-oriented, by pulsed laser deposition (PLD)–the growth conditions are reported elsewhere[36]–. Note that the use of single-crystal substrates of the same structure as the epitaxial film that is to be deposited (in this case the fluorite structure) facilitates coherent growth of the films and the minimization of grain boundaries or structure defects, and avoid the coexistence of different polymorphs which would act as electrically inactive phonon scattering centers, reducing the effectiveness of the $\kappa_{ON}$–$\kappa_{OFF}$ switching. In addition, YSZ intrinsically contains a large amount of oxygen vacancies and behaves as an oxide-ion conductor,[17,20,36] acting, hence, as a sink and source for the oxygen migration from/to the HZO film. The epitaxial strain exerted by the (111)-oriented YSZ substrates stabilizes the orthorhombic polar structure in the HZO films, with no traces of the non-polar monoclinic structure, as shown in the X-ray diffractograms around the 111 substrate reflection (see **Fig. S1** in the Supp. Info.). Moreover, the Laue fringes around the 111 film reflection (**Fig. S1**) indicate a high crystal quality of the films, as discussed in a previous work.[36] Non-reactive Au, 60 nm-thick pads, were ex-situ deposited on top of the HZO/YSZ samples by sputtering, acting as top electrodes and transducers for the frequency-domain thermoreflectance (FDTR) to measure the thermal properties under different applied voltages–see Methods and Supp. Info.–.[37–39] The thermal properties were obtained by fitting the frequency-dependent phase data, $\phi(f)$, of the FDTR experiments to an analytical solution of the heat diffusion equation in a multilayer model.[38] Due to the low thickness value of the HZO film (7 nm), the whole HZO layer is considered and modelled as a thermal boundary conductance (TBC, **Fig. 1a**) between the Au layer and the YSZ substrate (see further details in the Supp. Info.). The comparison between the raw phase-shift data recorded for Au/HZO/YSZ and Au/YSZ, shows that the insertion of the YSZ film results in an increase of the phase at intermediate frequencies (in the range of maximum sensitivity to the TBC) consistent with an increase of the thermal boundary resistance of the device (**Fig. 1b**). Also note that in the frequency range between $\approx 10^5$ Hz and $\approx 10^7$ Hz the effects of different thermal boundary resistances are maximized (**Fig. 1b**), corresponding to the frequency range where the sensitivity of the TBC is the largest (**Fig. 1c**). By determining the TBC between the Au film and YSZ in the Au/HZO/YSZ samples using the heat diffusion equation in a multilayer model,[38] the $\kappa$ of the HZO film is computed by multiplying the TBC by the thickness (see further details in the Supp. Info.). A similar approach was considered previously.[40]

Next, the FDTR measurements were performed at 200ºC on the Au/HZO/YSZ sample while applying different voltages from the top electrode (Au) and using YSZ as grounded bottom electrode (**Fig. 2a**). Some representative experimental $\phi(f)$ data (solid squares) at different applied voltages and the corresponding fitting (solid lines) to heat diffusion equation in a multilayer model are shown in **Fig. 2b**.[37,38] As observed, the thermal boundary resistance of the HZO layer significantly changes with the applied electric field, which manifests itself in the previously mentioned frequency range of the $\phi(f)$ data (between $\approx 10^5$ Hz and $\approx 10^7$ Hz). The resulting $\kappa$ values are shown in **Fig. 2c** and they vary from a minimum of $\approx 0.35$ Wm$^{-1}$K$^{-1}$ to a maximum $\approx 0.55$ Wm$^{-1}$K$^{-1}$ depending on the applied electric field. These values are slightly below the reported κ in 20 nm thick $Hf_{1-x}Zr_xO_2$ films (κ ≈ 0.7 – 1.2 Wm$^{-1}$K$^{-1}$),[40] due to the reduced thickness of our films. Indeed, for a hypothetic bulk single-crystal ferroelectric HZO sample a much larger κ is predicted (≈ 5 Wm$^{-1}$K$^{-1}$).[41]

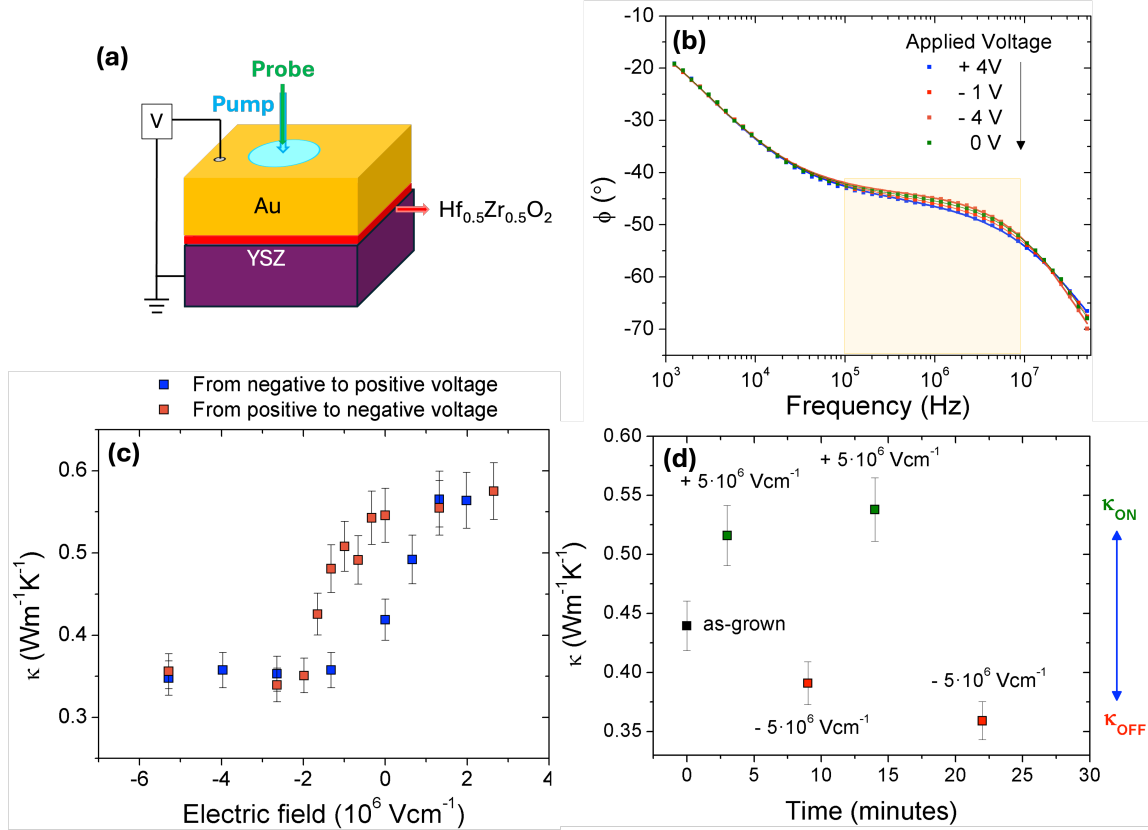

*Fig. 2. (a) Sketch of the FDTR measurements performed while applying voltages on the Au/Hf$_{0.5}$Zr$_{0.5}$O$_2$/YSZ samples at 200ºC. (b) Frequency dependent phase data of the FDTR measurements at different voltages. The frequency range where the effects of TBC are maximized is shadowed in yellow. (c) Electric-field dependence of the thermal conductivity of Hf$_{0.5}$Zr$_{0.5}$O$_2$ films at 200ºC. (d) κ$_{ON}$–κ$_{OFF}$ switching by applying positive and negative voltages.*

Strikingly, the electric-field dependence of κ in our HZO films is clearly hysteretic (Fig. 2c) and resembles the polarization-electric field hysteresis loop–with similar coercive field values–previously reported.[36] To the best of our knowledge, this is the first κ(E) hysteresis loop ever reported in a ferroelectric material. This result allows defining two non-volatile thermal conductivity states (κ$_{ON}$ and κ$_{OFF}$ for the high and low thermal conductivity states, respectively). Moreover, the κ$_{ON}$–κ$_{OFF}$ states are reproducible, allowing to recover the same thermal states after successive switching with the same +/- voltages (**Fig. 2d**), providing a κ$_{ON}$/κ$_{OFF}$ ratio ≈ 1.6.

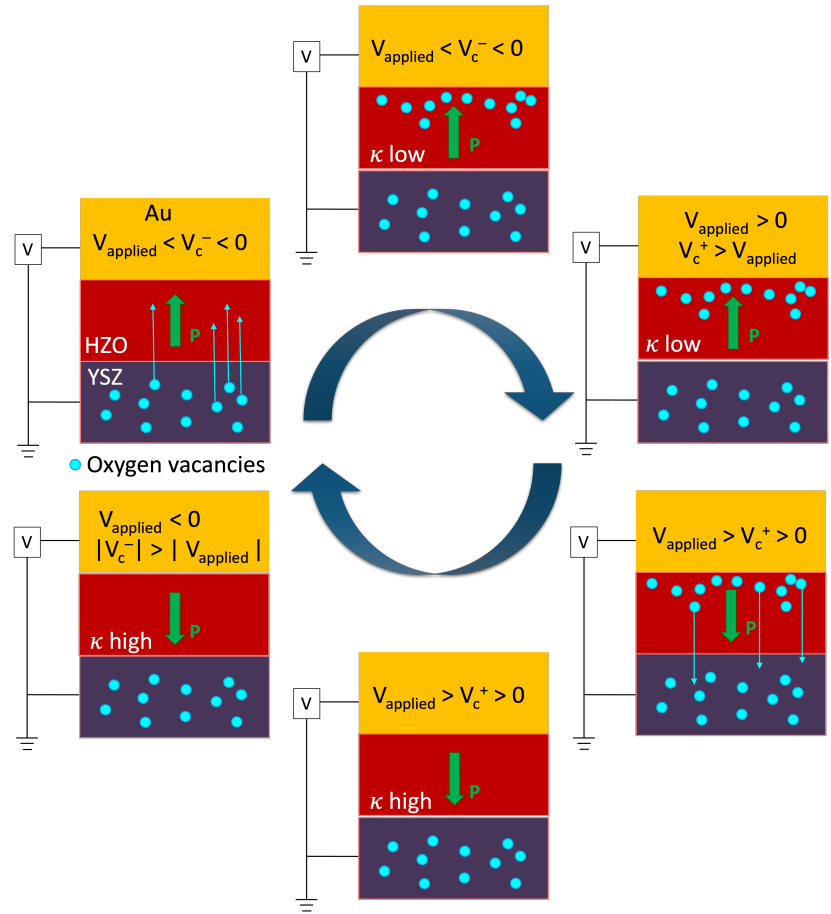

*Fig. 3*. *Sketch showing the coupling mechanism between oxygen vacancies migration and polarization in Hf$_{0.5}$Zr$_{0.5}$O$_2$ films and its effects on the κ(E) hysteresis loop.*

The κ(E) hysteresis loop shown in **Fig. 2c** is at odds with domain walls being the main phonon scattering mechanism reported in conventional ferroelectrics.[1,2,4,9] If that were the case, the κ(E) response should be quite different: the minimum κ should be around the coercive fields, when the density of domain walls is large; conversely, κ should be maximum at both large positive fields and large negative fields, when the ferroelectric is single domain and thus domain walls are absent. Here, instead, κ is minimum at large negative voltage and maximum at large positive voltage. Therefore, domain walls cannot be the main contributors to the observed κ(E) response in our ferroelectric HZO films.

Instead, the role of the oxygen vacancy migration with electric field previously reported for YSZ and HZO[17,20,35] and its coupling with the ferroelectric polarization of HZO can explain this thermal behavior. This explanation is sketched in **Fig. 3** and summarized as follows. Applying a large enough negative voltage (larger than the coercive voltage, $V_c$; see **Fig. 3**) to the top metallic electrode switches the out-of-plane component of the polarization of the HZO layer towards it; at the same time, the positively charged oxygen vacancies migrate from the YSZ reservoir through the YSZ/HZO interface. Note that the non-reactive Au top electrode ensures that the oxygen vacancies stay in the HZO layer, and the ferroelectric polarization of the oxide retains the vacancies against diffusion when the electric field is removed. Actually, the concentration of oxygen vacancies, and hence κ, is kept stable on further increasing the voltage towards positive values, until above $V_c^+$ the polarization switches towards the YSZ substrate. At this point, oxygen vacancies rapidly migrate towards the YSZ, recovering the κ of HZO again. Therefore, the coupling between the polarization and the oxygen vacancy migration allows the precise and hysteretic tuning of κ in hafnia-based epitaxial films. Although this coupling between polarization and oxygen ion migration was also proved in epitaxial ferroelectric $BaTiO_3$ films embedded between a topotactic oxide ($SrCoO_x$) and the ionic liquid,[42] and, to a certain extent, in $BaTiO_3$ epitaxial films grown on Nb:$SrTiO_3$,[43] it has never been used as a mechanism to electrically manipulate κ. Ionic control of the metal-oxide thermal boundary conductance in Pt/$SrTiO_3$/Nb:$SrTiO_3$ devices was recently reported, although in this case the absence of polarization results in spontaneous migration of oxygen vacancies against the chemical potential gradient and a gradual relaxation of the thermal states.[44]

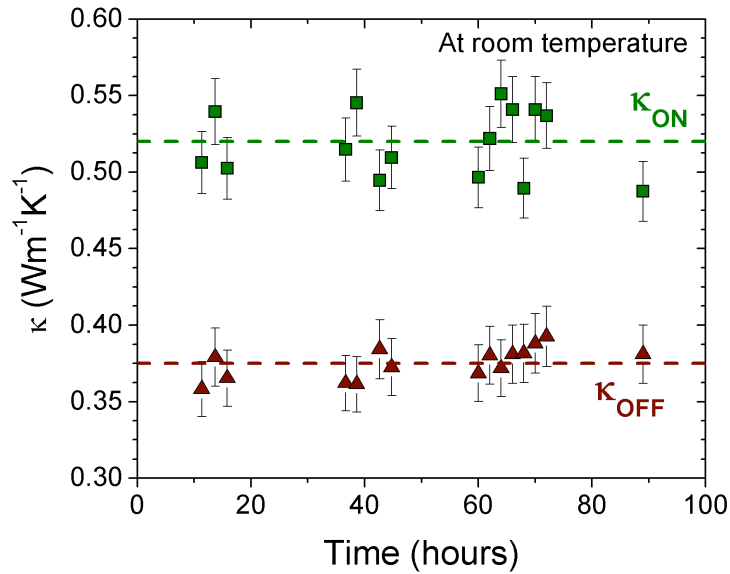

*Fig. 4*. Stability of the $κ_{ON}$ and $κ_{OFF}$ values at room temperature with no electric field applied.

To evaluate the robustness of the $\kappa_{ON}$ and $\kappa_{OFF}$ states, a HZO film was polarized under positive/negative electric fields and the thermal conductivity of the ON/OFF states was followed at room temperature as a function of time. As observed in **Fig. 4**, both states are stable for several days under laboratory conditions. This confirms that spontaneous diffusion of oxygen vacancies is strongly suppressed at room temperature, due to the polarization of the HZO film.

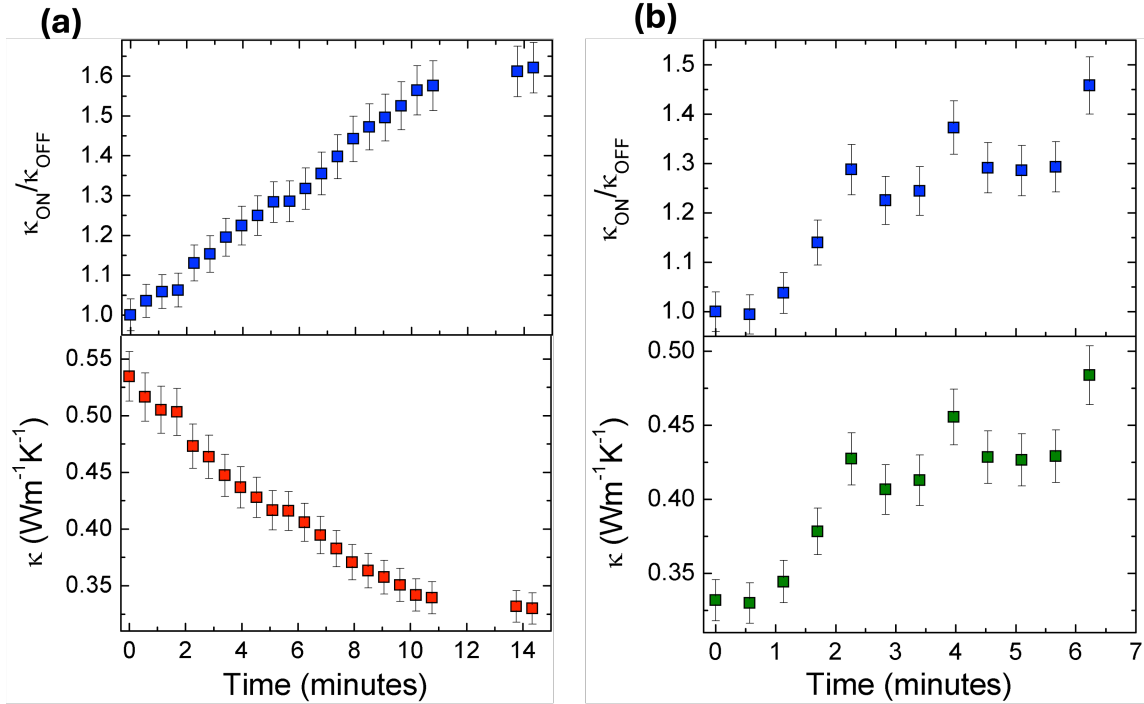

*Fig. 5*. $\kappa_{ON}/\kappa_{OFF}$ *ratio (top panels) and thermal conductivity (lower panels) as a function of time when immediately switch **(a)** from +3V to –3V and **(b)** from –3V to +3V at 200ºC.*

Regarding the speed of the process,[34] **Fig. 5** shows the evolution of the $\kappa$ of HZO film with time after switching from –3V to +3V and vice versa. As observed, it takes several minutes to reach the maximum $\kappa_{ON}/\kappa_{OFF}$ ratio, which signals the time required to reach the equilibrium population of oxygen vacancies between the film and the YSZ reservoir. A much larger oxide-ion mobility has been reported in HZO at room temperature (it must be even faster at 200ºC) when grown on $Nb:SrTiO_3$ substrates;[35] we thus conclude that the $\kappa_{ON}$–$\kappa_{OFF}$ switching time in our samples could be limited by the ion mobility at the YSZ/HZO interface or in the YSZ substrate itself. In fact, in a previous work a faster oxygen ion mobility in YSZ was achieved by increasing, not only the temperature (at 280ºC), but also the applied voltage (8V).[20] However, in the Au/HZO/YSZ samples, either the increase in temperature or the applied voltage resulted in a rapid deterioration of the Au/HZO interface, detaching the Au layer from the HZO film, which impeded the FDTR and the electrical measurements.

**Conclusions.**

In this work we demonstrate the potential of ferroelectric $Hf_{1-x}Zr_xO_2$ as a thermally tunable material for solid-state thermal memory applications. The findings reveal that the electric-field dependence of thermal conductivity in HZO films exhibits a hysteresis loop, resembling the polarization-electric field behavior, which enables the definition of non-volatile $\kappa_{ON}$ and $\kappa_{OFF}$ states. The coupling of oxygen vacancy migration with ferroelectric polarization was identified as the primary mechanism influencing thermal conductivity, offering an alternative to phonon scattering via electric-field-reconfigurable domain walls. Moreover, a $\kappa_{ON}/\kappa_{OFF}$ ratio of 1.6 was achieved, representing the highest value recorded for ferroelectric materials under electric field stimuli above room temperature. On the other hand, the $\kappa_{ON}$ and $\kappa_{OFF}$ thermal states proved to be highly stable, retaining their values over days, which suggests minimal oxygen vacancy rediffusion at room temperature. However, the $\kappa_{ON}$–$\kappa_{OFF}$ switching time was slower than anticipated, primarily due to limitations in ion mobility within the YSZ substrate, despite the rapid mobility previously reported in the HZO layer itself. Overall, this work on HZO films present a new viable approach for actively modulating the thermal conductivity, which can open a new path in the engineering of electrically controlled thermal memory devices.

**References**

1. Ihlefeld, J. F. *et al.* Room-temperature voltage tunable phonon thermal conductivity via reconfigurable interfaces in ferroelectric thin films. *Nano Lett* **15**, 1791–1795 (2015).
2. Foley, B. M. *et al.* Voltage-Controlled Bistable Thermal Conductivity in Suspended Ferroelectric Thin-Film Membranes. *ACS Appl Mater Interfaces* **10**, 25493–25501 (2018).
3. Aryana, K. *et al.* Observation of solid-state bidirectional thermal conductivity switching in antiferroelectric lead zirconate (PbZrO3). *Nat Commun* **13**, 1–9 (2022).
4. Negi, A. *et al.* Ferroelectric Domain Wall Engineering Enables Thermal Modulation in PMN–PT Single Crystals. *Advanced Materials* **35**, (2023).
5. Seijas-Bellido, J. A. *et al.* A phononic switch based on ferroelectric domain walls. *Phys Rev B* **96**, 140101(R) (2017).
6. Liu, C. *et al.* Large Thermal Conductivity Switching in Ferroelectrics by Electric Field-Triggered Crystal Symmetry Engineering. *ACS Appl Mater Interfaces* **14**, 46716–46725 (2022).

7. Wooten, B. L. *et al.* *Electric Field-Dependent Phonon Spectrum and Heat Conduction in Ferroelectrics*. https://www.science.org (2023).
8. Féger, L. *et al.* Lead-free room-temperature ferroelectric thermal conductivity switch using anisotropies in thermal conductivities. *Phys Rev Mater* **8**, (2024).
9. Lin, Y. *et al.* Ferroelectric Polarization Modulated Thermal Conductivity in Barium Titanate Ceramic. *ACS Appl Mater Interfaces* **14**, 49928–49936 (2022).
10. Huang, H. T., Lai, M. F., Hou, Y. F. & Wei, Z. H. Influence of magnetic domain walls and magnetic field on the thermal conductivity of magnetic nanowires. *Nano Lett* **15**, 2773–2779 (2015).
11. Pocs, C. A. *et al.* Giant thermal magnetoconductivity in $CrCl_3$ and a general model for spin-phonon scattering. *Phys Rev Res* **2**, 1–13 (2020).
12. Nakayama, H. *et al.* Above-room-temperature giant thermal conductivity switching in spintronic multilayers. *Appl Phys Lett* **118**, (2021).
13. Cazorla, C. & Rurali, R. Dynamical tuning of the thermal conductivity via magnetophononic effects. *Phys Rev B* **105**, (2022).
14. Shin, J. *et al.* Light-triggered thermal conductivity switching in azobenzene polymers. *Proc Natl Acad Sci U S A* **116**, 5973–5978 (2019).
15. Langenberg, E. *et al.* Ferroelectric Domain Walls in $PbTiO_3$ Are Effective Regulators of Heat Flow at Room Temperature. *Nano Lett* **19**, 7901–7907 (2019).
16. Lu, Q. *et al.* Bi-directional tuning of thermal transport in $SrCoO_x$ with electrochemically induced phase transitions. *Nat Mater* **19**, 655–662 (2020).
17. Yang, Q. *et al.* Solid-State Electrochemical Thermal Transistors. *Adv Funct Mater* **33**, (2023).
18. Zhang, Y. *et al.* Wide-range continuous tuning of the thermal conductivity of $La_{0.5}Sr_{0.5}CoO_{3-\delta}$ films via room-temperature ion-gel gating. *Nat Commun* **14**, (2023).
19. Li, H. B. *et al.* Wide-range thermal conductivity modulation based on protonated nickelate perovskite oxides. *Appl Phys Lett* **124**, (2024).
20. Varela-Domínguez, N., Claro, M. S., Carbó-Argibay, E., Magén, C. & Rivadulla, F. Exploring Topochemical Oxidation Reactions for Reversible Tuning of Thermal Conductivity in Perovskite Fe Oxides. *Chemistry of Materials* (2024) doi:10.1021/acs.chemmater.4c02023.
21. Wehmeyer, G., Yabuki, T., Monachon, C., Wu, J. & Dames, C. Thermal diodes, regulators, and switches: Physical mechanisms and potential applications. *Appl Phys Rev* **4**, (2017).
22. Swoboda, T., Klinar, K., Yalamarthy, A. S., Kitanovski, A. & Muñoz Rojo, M. Solid-State Thermal Control Devices. *Adv Electron Mater* **7**, (2021).
23. Liu, C. *et al.* Actively and reversibly controlling thermal conductivity in solid materials. *Physics Reports* vol. 1058 1–32 Preprint at https://doi.org/10.1016/j.physrep.2024.01.001 (2024).
24. Nataf, G. F. *et al.* Using oxides to compute with heat. *Nature Reviews Materials* vol. 9 530–531 Preprint at https://doi.org/10.1038/s41578-024-00690-1 (2024).
11

25. Forman, C., Muritala, I. K., Pardemann, R. & Meyer, B. Estimating the global waste heat potential. *Renewable and Sustainable Energy Reviews* **57**, 1568–1579 (2016).
26. Spaldin, N. A. & Ramesh, R. Advances in magnetoelectric multiferroics. *Nat Mater* **18**, 203–212 (2019).
27. Breckenfeld, E. *et al.* Effect of growth induced (non)stoichiometry on the structure, dielectric response, and thermal conductivity of SrTiO 3 thin films. *Chemistry of Materials* **24**, 331–337 (2012).
28. Brooks, C. M. *et al.* Tuning thermal conductivity in homoepitaxial $SrTiO_3$ films via defects. *Appl Phys Lett* **107**, 051902 (2015).
29. Sarantopoulos, A., Ong, W. L., Malen, J. A. & Rivadulla, F. Effect of epitaxial strain and vacancies on the ferroelectric-like response of CaTiO3 thin films. *Appl Phys Lett* **113**, 182902 (2018).
30. Bugallo, D. *et al.* Deconvolution of Phonon Scattering by Ferroelectric Domain Walls and Point Defects in a PbTiO 3 Thin Film Deposited in a Composition-Spread Geometry . *ACS Appl Mater Interfaces* **13**, 45679–45685 (2021).
31. Varela-Domínguez, N., Claro, M. S., Vázquez-Vázquez, C., López-Quintela, M. A. & Rivadulla, F. Electric-Field Control of the Local Thermal Conductivity in Charge Transfer Oxides. *Advanced Materials* (2024) doi:10.1002/adma.202413045.
32. Materano, M. *et al.* Influence of oxygen content on the structure and reliability of ferroelectric HfxZr1−xO2 layers. *ACS Appl Electron Mater* **2**, 3618–3626 (2020).
33. He, R., Wu, H., Liu, S., Liu, H. & Zhong, Z. Ferroelectric structural transition in hafnium oxide induced by charged oxygen vacancies. *Phys Rev B* **104**, (2021).
34. Ma, L. Y. & Liu, S. Structural Polymorphism Kinetics Promoted by Charged Oxygen Vacancies in HfO2. *Phys Rev Lett* **130**, (2023).
35. Nukala, P. *et al. Reversible Oxygen Migration and Phase Transitions in Hafnia-Based Ferroelectric Devices Downloaded From*. Science vol. 372 http://science.sciencemag.org/ (2021).
36. Barriuso, E. *et al.* Epitaxy-Driven Ferroelectric/Non-Ferroelectric Polymorph Selection in an All-Fluorite System. *Adv Electron Mater* **10**, (2024).
37. Schmidt, A. J., Cheaito, R. & Chiesa, M. A frequency-domain thermoreflectance method for the characterization of thermal properties. *Review of Scientific Instruments* **80**, 094901 (2009).
38. Kirsch, D. J. *et al.* An instrumentation guide to measuring thermal conductivity using frequency domain thermoreflectance (FDTR). *Rev Sci Instrum* **95**, (2024).
39. Malen, J. A. *et al.* Optical Measurement of Thermal Conductivity Using Fiber Aligned Frequency Domain Thermoreflectance. *J Heat Transfer* **133**, 081601 (2011).
40. Scott, E. A. *et al.* Thermal resistance and heat capacity in hafnium zirconium oxide (Hf1-xZrxO2) dielectrics and ferroelectric thin films. *Appl Phys Lett* **113**, (2018).
12

41. Zhang, S., Yi, S., Yang, J. Y., Liu, J. & Liu, L. First-principles study of thermal transport properties in ferroelectric HfO2 and related fluorite-structure ferroelectrics. *Physical Chemistry Chemical Physics* **25**, 17257–17263 (2023).
42. Gu, Y. *et al*. Oxygen-Valve Formed in Cobaltite-Based Heterostructures by Ionic Liquid and Ferroelectric Dual-Gating. *ACS Appl Mater Interfaces* **11**, 19584–19595 (2019).
43. Lü, W. *et al*. Multi-Nonvolatile State Resistive Switching Arising from Ferroelectricity and Oxygen Vacancy Migration. *Advanced Materials* **29**, (2017).
44. Álvarez-Martínez, V. *et al*. Interfacial Thermal Resistive Switching in (Pt,Cr)/SrTiO3 Devices. *ACS Appl Mater Interfaces* **16**, 15043–15049 (2024).

**Supplementary information:**

**The coupling of polarization and oxygen vacancy migration in ferroelectric $Hf_{0.5}Zr_{0.5}O_2$ thin films enables electrically controlled thermal memories above room temperature.**

Dídac Barneo[1,2], Rafael Ramos[3], Hugo Romero[4], Víctor Leborán[4], Noa Varela-Domínguez[3], José A. Pardo[4,5], Francisco Rivadulla[3], Eric Langenberg[1,2,*]

[1]Departament de Física de la Matèria Condensada, Universitat de Barcelona, 08028 Barcelona, Spain.

[2]Institut de Nanociència i Nanotecnologia ($IN^2UB$), Universitat de Barcelona, 08028 Barcelona, Spain.

[3]Centro Singular en Química Biolóxica e Materiais Moleculares (CiQUS), Universidade de Santiago de Compostela, 15782 Santiago de Compostela, Spain.

[4]Instituto de Nanociencia y Materiales de Aragón (INMA), CSIC-Universidad de Zaragoza, 50009 Zaragoza, Spain.

[5]Departamento de Ciencia y Tecnología de Materiales y Fluidos, Universidad de Zaragoza, 50018 Zaragoza, Spain.

**Pulsed Laser Deposition of the films and X-Ray Diffraction characterization.**

$Hf_{0.5}Zr_{0.5}O_2$ (HZO) thin films were grown on 111-$Y_2O_3$:$ZrO_2$ (YSZ) by pulsed laser deposition (PLD). The substrate temperature was fixed at 800 ºC, the oxygen pressure during growth was set at 100 mTorr, and 78 mJ per pulse at a rate of 10 Hz was used. They were cooled down without any further change in oxygen pressure nor annealing.

The structure of the films was characterized by x-ray diffraction (XRD), showing that 111-oriented YSZ yields non-monoclinic phases of HZO (shoulder on the right of the peak of the 111-YSZ substrate, at around $2\theta \approx 30.5º$) as shown in Fig. S1. More details can be found in previous work.[1] X-ray reflectivity (XRR) was used to determine the thickness for the HZO films.

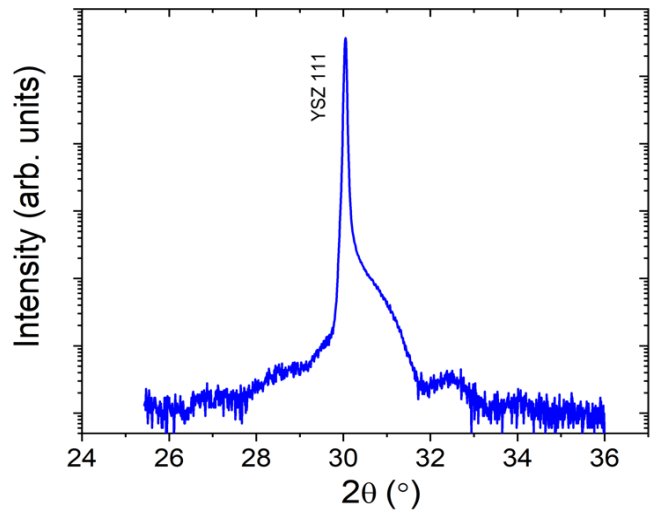

*Fig. S1.* X-ray diffraction pattern of HZO films deposited onto 111-oriented YSZ substrates.

**FDTR measurements.**

Thermal characterization was done by Frequency-Domain Thermoreflectance (FDTR).[2,3] FDTR is a non-contact optical pump–probe technique. It fundamentally works with two different laser beams: the first beam of light (the pump) acts as the heat source while the second one (the probe) detects the temperature change caused by the pump and the thermal dissipation ($\Delta T$) through a change in surface reflectivity ($\Delta R$): $\Delta T = (dR/dT)^{-1}\Delta R = (\beta)^{-1}\Delta R$. In order to quantify properly this reflectance (and its changes), a transducer is normally deposited on top of the studied materials, normally Au, and in our system, it was used as top electrode as well. Measurements are performed in a wide frequency range at constant irradiation, making a sweep in frequency, differently from Time-Domain Thermoreflectance where the variable magnitude is the time of the pulse from the laser (period). The variable heat source produces temperature gradients and transients which enables the measurement of the thermal boundary conductance (TBC) between Au and YSZ (or the equivalent HZO film as explained later) with good precision.

In our setup (Fig. S2a), a sinusoidal modulated pump laser ($\lambda$=405 nm, modulating frequency 2 kHz–50 MHz, Gaussian spot size $1/e^2$ radius ≈ 10.5 μm) is focused on the surface of the film, coated with a 60-nm-thick layer of Au. This sinusoidal irradiation produces an oscillatory modulation of the temperature in the surface, and therefore a periodic change of the Au thermoreflectance. The laser beam used to probe these phenomena (λ=532 nm) is split to measure both the signal before reaching the sample (reference) and after, getting a variation in the phase between both parts of the beam. Also, this process is used to improve the signal-to-noise ratio, especially at low frequencies, and calibrate well the phase-shift offsets produced by electronics and differences between optic paths.

Then, this phase shift measured in the stablised frequency range is fitted to an analytical solution of the heat diffusion equation for layered structures to determine the total thermal resistance ($R_{Tot}$).[3] More specifically, a multilayer model consisting on an Au film on top of an YSZ film is used, and the κ of the HZO film is obtained by considering it a thermal resistance, included in the TBC between Au and YSZ. This is done in this way because the HZO films are very thin compared to the other two materials, so the variations they produce in measurements are sufficiently small to be simplified for less complex (and more realistic) mathematical analysis. The model used is described in detail below (Fig. S2b).

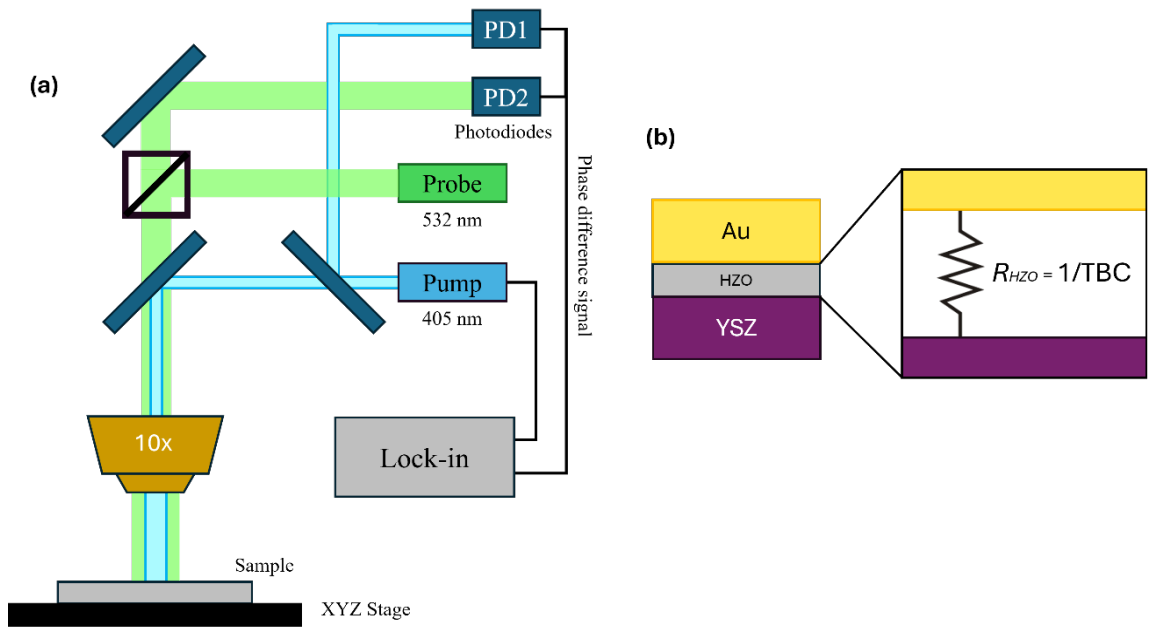

***Fig. S2.*** *Simplified schemes of **(a)** the FDTR setup and **(b)** the multilayer model used to fit the FDTR data, considering the HZO film as a thermal resistance.*

The specific heat capacity values, Cp, used for the fitting were extracted from the literature for YSZ and Au,[4,5] as well as the thermal conductivity of YSZ.[6] κ of Au was obtained by measuring its electrical conductivity, and applying the Wiedemann-Frantz law to obtain its thermal counterpart. YSZ thickness is taken from the nominal value (0.5 mm) provided by the supplier (Crystec), and the thickness of the Au transducer layer was quantified by XRR. Each parameter has different sensitivities upon frequency range.

*Table S1. Parameters used for fitting the FDTR data*

|  | Cp (MJ·K$^{-1}$·m$^{-3}$) | κ⊥ (W·m$^{-1}$·K$^{-1}$) | Thickness (nm) | TBC (W·m$^{-2}$·K$^{-1}$) |
|---|---|---|---|---|
| Au transducer | 2.2 | 90 | 60 | Fitting Parameter |
| YSZ substrate | 2.8 | 2.5 | 50,000 | |

**References**

(1) Barriuso, E.; Jiménez, R.; Langenberg, E.; Koutsogiannis, P.; Larrea, Á.; Varela, M.; Magén, C.; Algarabel, P. A.; Algueró, M.; Pardo, J. A. Epitaxy-Driven Ferroelectric/Non-Ferroelectric Polymorph Selection in an All-Fluorite System. *Adv Electron Mater* **2024**, *10* (5). https://doi.org/10.1002/aelm.202300522.

(2) Schmidt, A. J.; Cheaito, R.; Chiesa, M. A Frequency-Domain Thermoreflectance Method for the Characterization of Thermal Properties. *Review of Scientific Instruments* **2009**, *80*, 094901. https://doi.org/10.1063/1.3212673.

(3) Kirsch, D. J.; Martin, J.; Warzoha, R.; McLean, M.; Windover, D.; Takeuchi, I. An Instrumentation Guide to Measuring Thermal Conductivity Using Frequency Domain Thermoreflectance (FDTR). *Rev Sci Instrum* **2024**, *95* (10). https://doi.org/10.1063/5.0213738.

(4) Khvan, A. V.; Uspenskaya, I. A.; Aristova, N. M.; Chen, Q.; Trimarchi, G.; Konstantinova, N. M.; Dinsdale, A. T. Description of the Thermodynamic Properties of Pure Gold in the Solid and Liquid States from 0 K. *CALPHAD* **2020**, *68*. https://doi.org/10.1016/j.calphad.2019.101724.

(5) Salamatov, E. I.; Taranov, A. V.; Khazanov, E. N.; Charnaya, E. V.; Shevchenko, E. V. Transport Characteristics of Phonons and the Specific Heat of Y2O3:ZrO2 Solid Solution Single Crystals. *Journal of Experimental and Theoretical Physics* **2017**, *125* (5), 768–774. https://doi.org/10.1134/S1063776117100144.

(6) Langenberg, E.; Ferreiro-Vila, E.; Leborán, V.; Fumega, A. O.; Pardo, V.; Rivadulla, F. Analysis of the Temperature Dependence of the Thermal

Conductivity of Insulating Single Crystal Oxides. *APL Mater* **2016**, *4*, 104815. https://doi.org/10.1063/1.4966220.